\documentclass[12pt,preprint]{aastex}
\usepackage{graphicx}

\input{psfig}

\shorttitle{Misaligned disks in halos}
\shortauthors{Quadri, M\"oller \& Natarajan}

\begin{document}

\title{Lensing effects of misaligned disks in dark matter halos}

\author{Ryan Quadri$^{1}$, Ole  M\"oller$^{2}$ \& Priyamvada Natarajan$^1$}
\affil{1 Department of Astronomy, Yale University, 260 Whitney Avenue, 
New Haven, CT 06511, USA}
\affil{2  Kapteyn Institute, PO Box 800, 9700 AV Groningen, The Netherlands}

\begin{abstract}
In this paper we study the observational signatures of the lensing
signal produced by dark matter halos with embedded misaligned
disks. This issue is of particular interest at the present time since
most of the observed multiple lens systems have magnification ratios
and image geometries that are not well-fit by standard mass
models. The presence of substructure exterior to the lens has been
invoked by several authors in the context of Cold Dark Matter (CDM)
models in order to explain the anamolous magnification ratios.  We
emphasize that the anomalous magnification ratios may be an artifact
of the simple one-component mass models currently in use; the
inclusion of a misaligned disk may be able to mimic the effect of
substructure.  These slight spatial offsets between the dark matter
halo and the disk, which are likely to occur during or as a
consequence of interactions or mergers, lead to complex image
configurations and non-standard magnification ratios.  We investigate
the effects of disk misalignment on two illustrative lenses: a spiral
disk embedded within a dark matter halo, and a compact disk-like
component within an elliptical galaxy. The expected fraction of
galaxies with a misaligned disk is estimated to be of the order of
10\%. In such cases we find that the resultant lensing geometries are
unusual, with high image multiplicities. The caustic structures - both
radial and tangential - are drastically modified and the magnification
ratios differ compared to expectations from standard lens models. The
additional parameters required to specify the relative alignment of
multiple mass components in the primary lens introduce yet another
source of uncertainty in the mass modeling of gravitational lens
systems.
\end{abstract}

\keywords{gravitational lensing, galaxies: fundamental parameters, 
halos, methods: numerical}

\newpage

\section{Introduction}

Lensing by individual galaxies provides a plethora of observational
signatures ranging from multiply imaged, highly magnified background
sources to a weakly sheared background (see Schneider, Ehlers \& Falco
1992, Blandford \& Narayan 1992). Several authors have attempted to
use these observed image geometries, positions and magnitudes to infer
and reconstruct the mass profile of the lens using ground-based
optical and radio surveys (Browne et al. 2001 [JVAS/CLASS] and
references therein; Lehar et al. 2000 [CASTLES]). Attempts have been
made to constrain both the density profiles of the mass distribution
within the inner regions of galaxies (Keeton \& Madau 2001; Cohn et
al. 2001) as well as obtain constraints on the Hubble parameter $H_0$
(Kundi{\'c} et al. 1997; Fassnacht et al. 2002). Both of these
enterprises are complicated by the presence and precise configuration
of structure within and exterior to the lensing galaxy in
question. The degeneracies in such modeling attempts arise due to the
presence of structural complexity in the inner regions of galaxies,
the prevalence of disky components, bulges, central black holes
(M\"oller \& Blain 1998; Mao, Witt \& Koopmans 2001), as well as the
presence of external mass perturbations: nearby galaxies, groups or
clusters in the vicinity of the lens (Keeton, Kochanek \& Seljak 1997;
M\"oller, Natarajan, Kneib \& Blain 2002). In this work, we explore a
further source of uncertainty - the presence of misaligned disks within 
dark halos.

The motivation for studying galaxy scale systems that have their
baryonic disks misaligned slightly with their dark matter halos is
two-fold: first, given what is currently accepted regarding the
heirarchical assembly of structure in cold dark matter dominated
structure formation models (Frenk \& White 1991) i.e. that merging of
galaxies is frequent - then dynamically induced small
misalignments between the disk and halo are expected during the merger
process. Since the lensing properties of a system (an individual
galaxy in this case) depend on the projected surface mass density, we
demonstrate in the following sections that these small misalignments
are likely to have very interesting and important observable
consequences. Secondly, it has been recently claimed in the literature
(Schneider \& Mao 1998; Metcalf \& Madau 2000; Dalal \&
Kochanek 2002) that the observed magnification ratios of several if
not most multiply imaged systems (notably B1422+231 and PG1115+08) can
only be explained as arising due to the presence of small scale
substructure within galaxy halos. Here, we illustrate that there are
in fact other physical scenarios such as an anisotropic mass
distribution within the lensing galaxy: non-zero quadrupole and
octopole components induced in the lensing potential due to a
misaligned disky component can also give rise to magnification
ratios and image characteristics that differ from those predicted by
the standard isothermal/pseudo isothermal mass models currently in
use. While the evidence for substructure is compelling it is useful to
explore the degeneracy with primary lens properties.

\section{When are halos and disks misaligned?}

In the standard paradigm for structure formation in cold dark matter
dominated models, galaxies are assembled over time via the merger of
smaller sub-clumps (Frenk \& White 1991). The merger rate for halos 
of a given mass can be computed semi-analytically using the extended 
Press-Schechter formalism (see Cole et al. 1995 and subsequent
refinements by Sheth, Mo \& Tormen 2001), and can be directly 
compared with cosmological N-body simulations (Raig et al. 2001). When 
dark matter halos merge their baryonic components also
merge; detailed numerical simulations of this process have been
carried out by Mihos \& Hernquist (1994) and Barnes et al. (1998). In
these simulations it is found that the baryonic gas flows to the
center of the more massive halo and the inner regions of the halo
become baryon-dominated. The tidal torquing during the merger process
is significant and can easily cause the embedded disk and halo to be
misaligned.  Misalignments of less than an arcsecond are highly
probable. It is also likely that during the `relaxation' process the
disk wanders transiently within the halo (Nelson \& Tremaine 1995;
Dubinski \& Kuijken 1995). Finally, interactions that do not result
in mergers can also perturb the disk-halo configuration  and induce
transient misalignments. These are  expected to be damped well within
a Hubble time (Dubinski \& Kuijken 1995). Since massive galaxies
are assembled via mergers, and because such galaxies are particularly
efficient as lenses, it is likely that even transitory 
misalignments could be important for a number of lens systems.

\subsection{Estimating the number of misaligned disks in halos}

Lens galaxies tend to sample the high mass end of the galaxy mass
function. As claimed above, such systems are expected to be built up
by mergers.  Misalignments are highly probable during this stage and
are likely to persist post-merger. We use the observationally
determined galaxy merger rate to estimate the expected frequency of
misaligned systems.  Patton et al. (2002) report a merger rate of
$\frac{dN}{dz} \propto (1 + z)^{2.3\pm 0.7}$ from a study of a sample
of bright galaxies, $-18 < M_B <-21$, in the CNOC2 redshift survey. We
assume here that all mergers lead to misalignments due to tidal
torquing during the merger process and that these misalignments
persist for the duration of the merger. From detailed N-body
simulations of interacting galaxies, the merger time-scale is found to
be of order 1 Gyr (see Fig.~4 of Mihos \& Hernquist
1994). Since typical lenses lie within a redshift range $z = 0.1 -
0.5$, the misaligned fraction is roughly,

\begin{eqnarray}  
f_{\rm misalign} \sim (\frac{t_{\rm merger}}{\delta t})\,
\,\frac{\it V(0.1,0.5)}{\it V(0,5)}\,\,
\frac{\int_{z=0.1}^{z=0.5}(\frac{dN}{dz})\,dz}{{\int_{z=0.1}^{z=0.5}}
\,dN\,}\sim 10\%,
\end{eqnarray}
where ${\it V (z_1,z_2)}$ is the comoving volume between redshifts $z_1$ and
$z_2$; $t_{\rm merger} \approx 1\,\mathrm{Gyr} $ is the merger timescale, and $\delta t$ the total elapsed time from 
$z = 0.1\,-\,0.5$. With the assumptions given above, $f_{\rm misalign}$
is found to be about 10\% in a $h=0.6$, $\Omega_{\lambda}=0.7$, $\Omega_{\mathrm{m}}=0.3$ cosmology. Note that the merger rate found by Patton et al. 
(2002) is consistent with semi-analytic estimates from models that are 
calibrated to observational data, as shown in the top panels of Fig.~6 in 
Somerville et al. (2000).

At present, there are approximately 50 confirmed cases of
individual galaxies lensing high redshift background quasars or
galaxies. We expect that at least a few of these are likely to
contain misaligned disks. Note that while the majority of lenses
detected thus far are typically early-type galaxies, observationally
it is becoming clear that early-type galaxies also do often contain
embedded compact disky components (Rest et al. 2001).  In fact, we
demonstrate below that even a highly compact misaligned disk that lies interior
to the Einstein radius can produce the anomalous image
geometries. Therefore, our analysis is generic and applicable to all
lensing galaxies that contain a disky mass component.
   
\section{Methodology: The Ray-tracing method}

In order to study the strong lensing properties of a misaligned halo
and disk we used the multiple plane ray-tracing code developed by
M\"oller \& Blain (1998). Computing the lensing properties of
multi-component lens systems is complicated and can be very
time-consuming if this is not done with an optimized algorithm. The
code we employ here is based on the `grid search method' (Schneider,
Ehlers \& Flaco 1992) and uses an adaptive grid that is optimized to
calculate deflection angles for parametric multiple component lens
models quickly and accurately. A regular grid of triangle pairs on the
image plane is mapped to the source plane. The magnification at each
point on the source plane is calculated as the sum of the ratios of
the area of the unmapped triangles on the image plane to that of the
mapped triangles on the source plane.  Similarly, shears are
calculated from the distortions of mapped (source plane) and unmapped
(image plane) triangles. In order to reduce the computation time the
mapping process is iterated adaptively. The numerical inaccuracies on
magnifications and shears are very small, usually less
than 0.1\%. The numerical uncertainty when calculating the number of
images for a given point source is greater; faint images near a caustic
line may be missed by this method (Schneider, Ehlers \& Falco
1992).  However, increasing the resolution of the grid enough can
overcome this problem, and it does not affect any of the results
discussed in this paper.
   
The illustrative single lens systems studied here consists of a dark
halo and an embedded disk.  The halo model is a cored isothermal
sphere (CIS). We have have also investigated pseudo-isothermal
elliptical mass distributions (PIEMDs) with an ellipticity of
$\epsilon = 0.3$ (Kassiola \& Kovner 1993) and find that finite
ellipticity in the halo does not significantly alter our results. The
two parameters required to specify the CIS are the core-radius and the
mass within a radius $R$. The former is not well-determined by observations of real
galaxies. In order to test the effects of our choice of $r_{\rm c}$ on
the results we explore a wide range of values, including $r_{\rm
c}=0$.  The total halo mass contained within a 20 $\mathrm{kpc}$
radius was held constant for all the models considered here to enable
a sensible comparison.  The primary effect of varying $r_{\rm c}$ is
a change in the relative significance of the halo in the lensing system,
which affects the size of the elliptical caustic (see \S~4.1).  This
is due to the fact that a small core radius implies that the
convergence $\kappa$ in the central region is very high. In this case,
the mass profile is centrally concentrated and thus the cross section
for strong lensing is significantly increased.

\section{Lens properties of simulated spiral systems}

We take the Milky Way as the template for our spiral lens,
and hence model the halo as a CIS with a small core radius $r_{\rm c}
= 0.5\,\mathrm{kpc}$ and mass $M_{\rm h} = 3.64 \times 10^{11}\,M_\sun$ within
a radius of 20\,$\mathrm{kpc}$.  This corresponds roughly to a velocity
dispersion of 160 km/s and the Einstein radius of the halo is ~0.4''.
The disk has mass $M_{\rm d} = 6.4 \times 10^{10}\,M_\sun$ and
scale length $r_{\rm d} = 3.2\,\mathrm{kpc}$.  Note that models with less
massive and more compact disks are also studied and the results are
presented in a subsequent section. An inclination angle of 70 degrees
was assumed for the disk which is roughly the minimal inclination
needed to significantly increase the cross-section for multiple
imaging (see Fig.~2 in M\"oller \& Blain 1998). The lens is located
at $z = 0.2$ and the source plane at $z = 1.5$. The values assumed for
cosmological parameters are $H_0=60 {\rm\, km\,s^{-1}\,Mpc^{-1}}$,
$\Omega_{\rm m}=0.3$, and $\Omega_{\lambda}=0.7$.  We have also
experimented with alternate cosmologies, such as $\Omega_{\rm
m}=1.0\,,\Omega_{\lambda}=0.0$ (EdS) and $\Omega_{\rm m}=0.3\,,
\Omega_{\lambda}=0.0$ (OCDM), and find that our results are not
significantly affected by the particular choice of these parameters.

In this paper we study simple spatial offsets of the disk relative to
the halo.  We assume that the disk retains an exponential profile and
that the halo retains an isothermal profile. Offsets in the plane of the
disk are denoted by $\Delta x$ and offsets outside the plane of the disk
are denoted $\Delta y$. The following sections assume 
offsets of $\Delta x=0.6''$, $\Delta y=0.6''$, and
$\Delta x=\Delta y=0.6''$.  We consider this to be a reasonable
approximation to the disturbed and asymmetric mass distribution that
will result from mergers or strong interactions.

\subsection{Magnification maps, cross sections and image geometries}

In this section we present magnification maps in the source plane for
several lens configurations, varying the position of the disk relative
to the halo in each case, i.e. the extent of spatial misalignment. We
also generate the corresponding magnification maps in the image plane,
where the value of each pixel is the magnification of the observed
image at that location. The caustics (points of infinite
magnification) in the source plane map onto critical curves in the
image plane. Magnification cross sections and representative image
geometries are also presented.

Fig.~1 shows the magnification maps in the source plane for various
representative values of the disk offset.  It is evident that a
realistic shift in the disk position relative to the halo does indeed
distort the caustic shapes severely. A shift in the x-direction, or in
the plane of the disk, yields a shift of the diamond-shaped curves
that form the \emph{tangential} caustic, which is caused by the
presence of the disk. The elliptical (\emph{radial}) caustic, which is
due to the peak in the central mass concentration of the combined halo
and disk, reduces in size and is distorted in the direction of the
offset. Offsets outside the plane of the disk lead to more complicated
caustic geometries. The tangential caustic shifts and the upper cusp
of the caustic appears to fold over into the diamond itself. The
radial caustic changes its shape drastically, and is pinched inward
near the lower cusp. But a comparison of Panel (c) and Panel (d) of
Fig.~1 shows that an offset in the x-direction can partially reduce
the effect of a shift in the y-direction.

Fig.~2 shows the magnification maps in the image plane. The small
inner critical curve corresponds to the radial caustic and the outer
one corresponds to the tangential, diamond-shaped caustic.  It is
clear from these images that the locations of highly magnified images
can be substantially affected by misalignments. As with the
magnification maps in the source plane, offsets out of the plane of
the disk also create complicated geometries. In Panel (c), Fig.~2, the
upper, roughly triangular shaped critical curve has split from the
inner critical curve. In Panel (d), Fig.~2, the region within this new
critical curve has nearly collapsed. If the disk were shifted further
in the x-direction then this new critical curve would be absorbed into
the outer critical curve. The evolution of caustics and critical
curves is continuous, i.e. the curves gradually distort, pinch off and
absorb each other.

To investigate the effects of misalignment on total magnification, we
calculate the magnification cross section ratio of the misaligned
cases to the aligned case.  The magnification cross section for a lens
is derived by adding up the total number of pixels with a value
greater than some threshold value $A$ in a magnification map.  The
results are shown in Fig.~3. An offset in the plane of the disk is
seen to have little effect. But the cross section for high
magnification $(A\,>\,30)$ events is significantly increased for
offsets outside the plane of the disk.  This is a manifestation of the
large surface area very near the caustic curves in Panel (c) of Fig.~2.
The offset in the y-direction of 0.6" corresponds approximately to the
maximum possible enhancement in high magnification cross section for
our lens model. The observed trend that offsets in the x-direction
tend to compensate for the more dramatic effects of shifts in the
y-direction is apparent here also.

This effect on the cross section for high magnifications is likely to
have a measurable effect on the statistics of multiple image
systems. Due to the preferential selection of high luminosity images
in a survey, any flux limited sample of lens systems is biased towards
systems with a large cross section for high magnifications ({\it
magnification bias}, e.g. Borgeest, 1991). Therefore the fraction of
lens systems with misaligned disks as estimated in \S~2.1
is conservative, and the actual number of systems in which a
misaligned disk is important might be larger.

The number of images of a source will increase or decrease by two if
the source crosses a caustic (see Schneider, Ehlers \& Falco 1992 for a 
mathematical illustration of the same).  An aligned halo and disk are 
capable of producing at most five separate images, in the configuration of a
four-image, possibly asymmetric, Einstein cross with a demagnified
central image.  However, our distorted caustic geometries for the
misaligned cases have regions where even higher image numbers are
possible. For instance, the lens configuration of Panel (c), Fig.~1 has
a region where seven images are formed and other lens configurations
can produce even more images.  Such a configuration with a large
multiplicity of images has been observed; B1359+154 is a 6-image system
(Rusin et al. 2001) where the lens is a compact group of galaxies at
$z = 1$.  These asymmetries in caustic geometries are expected to
yield complicated, asymmetric image geometries. Fig.~4 illustrates 
examples of the kind of image geometries that can be produced with the 
increased parameter space provided by disk offsets. The inset within each 
panel shows the magnification maps in the source plane and location of 
the point source.  A representative range of image geometries with a 
wide range of magnification ratios is shown. Current observations of such
systems might not reveal their complexity entirely; for example the 
configuration shown in Fig.~4, Panel (c) might be easily confused for 
an Einstein cross because the separate demagnified image is likely 
to go undetected. 

\section{Lens properties of early-type systems}

In the previous section we have concentrated on misalignment between
the disk and halo component in late-type lensing galaxies with Milky
Way-type spiral disks. The majority of lens systems known are, in
fact, early type galaxies, as these tend to be more massive and hence
have a larger cross section for lensing. Recent observations show that
disks or disk-like structures might well be present in early-type
lenses (Kelson et al. 2000). This suggests that early-type galaxies
may contain disks that may also be misaligned with respect to the
bulge/halo. Whether there is a possible misalignment between a disk
and the (baryonic) bulge depends on the exact cause of the
misalignment, but a scenario as described in \S~2 may cause a
misalignment between disk and the bulge as well as with the dark
matter halo.

\subsection{Magnification maps, cross-sections and image geometries}

We investigate here the lensing properties of a misaligned disk
component that is similar to that observed in ellipticals. In
particular, we choose disk parameters that are comparable to those
found by fitting the light profile of ellipticals in CL 1358+62 with a
bulge+disk model as in Kelson et al. (2000).  For our early-type lens
model, we choose a disk scale length of $r_{\rm d}=0.5\,{\rm kpc}$
which corresponds to the average disk scale length of the ellipitcals
in CL 1358+62 at the cluster redshift of $z=0.33$. The disk mass is
taken to be $M_{\rm h} = 2 \times 10^{10}\,M_\sun$.  We model the
bulge/halo as a CIS with $r_{\rm c} = 0.5\,\mathrm{kpc}$ and increase
the total mass within $20\,\mathrm{kpc}$ to $M_{\rm h} = 7 \times
10^{11}\,M_\sun$, to reflect the fact that early-type lens galaxies
are usually more massive. This choice of parameters gives a velocity
dispersion of ~220\,km/s and yields an Einstein radius of ~1'' for the
bulge/halo.

Fig.~5 shows the magnification maps in the source plane for the
elliptical lens.  The values for disk offsets are the same as in the
previous section.  Although this disk is less massive than the disk in
our late-type lens, it is also significantly more compact, meaning
that the central $\kappa$ value is in fact
high enough for the disk to produce an additional elliptical component
to the overall caustic structure.  So the magnification maps in Fig.~5
represent a superposition of two elliptical components due to the disk
and the halo, as well as the tangential caustic due to the disk. Fig.~6
shows the corresponding magnification map in the image plane. 
%Note
%that the maps have additional structure within the outer critical
%curve, including highly de-magnified regions, which cannot be seen
%clearly in the grayscale plot.  
As seen in Panel (a) of Figs.~5 and 6,
our model of a compact disk has virtually no effect on the structure
of the caustics or critical curves when the disk and halo are aligned.
The circular critical curve in Fig.~6 is nearly a circular Einstein
ring.  The other panels of Fig.~6 show critical curves that are
introduced by a misalignment of the compact disk.

Even though the disk components in the elliptical lens galaxy model
contain a smaller fraction of the total mass than the disks in the
late-type lens models discussed in previous sections, the effect of
the misalignment between disk and halo is still very pronounced. In
particular for misalignments in both the x and y direction, the effect
on the structure of the caustics and critical curves is drastic. Since
the majority of observed 4-image lens systems are expected to be
produced by lensing of a point source that lies close to the
high-magnification caustic lines, this indicates that the image
properties of these systems are affected strongly by the presence of a
misaligned disk, even if it is contributes only a small fratction of
the total mass and has a scale length of less than a few hundred parsecs.

It is therefore likely that the image geometries will also change
significantly. In Fig.~7 we show some typical image geometries of
early-type lens systems with a misaligned disk for two different lens
models.  The boxes are centered on the image positions and have sizes
that are proportional to the logarithm of the magnification. The
numbers near the images give the magnifications of the image.
Depending on the source position, both common and unusual image
geometries may be produced. The bottom-left panel shows a fairly
typical 3-1 geometry produced by a source near a cusp.  In contrast,
the top-left panel shows a case with non-standard magnification
ratios; the central image is not the brightest image.  These
`inverted' magnification ratios are a unique feature of complex
multi-component lens models; as discussed below, single component mass
models predict that the central image of the group of three images in
3-1 lens geometries is always the brightest.  The top-right panel of
Fig.~7 shows a lensed source near a cusp, but the 3-1 image geometry
is not present.  Instead there are four clustered images, with the two
brightest ones nearly overlapping, and a fifth image lying at a
distance of a few arcseconds.  Finally, the source in the bottom-right
panel does not lie on a cusp and the 3-1 image geometry is broken.
All images have roughly similar magnification and there is no clear
group of three closely grouped images.

\section{Interacting halos}

We point out in this section that even grazing interactions between
neighboring halos are likely to produce transient misalignments due to
tidal torquing. While these offsets might not persist for significant
periods of time, they are worth exploring as a special case of the
generic misaligned case. We use the disk$+$halo model described in
\S~4 but add a secondary halo with about 30\% of the mass of the
primary at a distance of one to several arc-seconds away from the
primary lens. Simulations of such interacting halos suggest that the
disk misalignment will tend to be along the direction of the
secondary. We have explored a variety of halo separations and choose a
distance of 1.7 arc-seconds to illustrate the effect. This separation
in fact corresponds to a spatial distance of 2$r_{\rm d}$, where
$r_{\rm d} = 3.2\,$ kpc is the disk scale length for the proto-type
Milky Way type spiral. The magnification maps in the source plane are
shown in Fig.~8. The main effects are an increase in the area enclosed
by the caustics, an overall shift of the caustic structure, and
induced distortions in the direction of the secondary halo. Comparing
Panel (c) of Fig.~1 to Panel (c) of Fig.~8 also shows that the radial
caustic is less pinched at the lower edge.  Close pairs of galaxies
are in fact observed as lenses; B1608+656 (Koopmans \& Fassnacht 1999)
is a clear example of this. If one or both of the lens galaxies in
this system contain discs, a scenario as discussed above is likely to
arise.

\section{Magnification ratios}

There has been much interest recently in explaining some of the
observed magnification ratios with lens models that include compact
substructure with masses that are approximately $10^{6}
M_{\odot}$. Such substructure is ubiquitous in numerical simulations
of structure formation in standard cold dark matter cosmologies, but
has not yet been observed. An argument in favor for the possible
ubiquity of dark clumps is that there are many viable mechanisms that
could prevent star formation in such compact sub-halos (Somerville
2002) so that they could indeed exist but be unobservable directly.

Gravitational lensing does in principle provide a method to detect
dark matter substructure on these small scales, and the recent
arguments presented by Metcalf \& Madau (2001), Metcalf \& Zhao
(2002), Dalal \& Kochanek (2002) and Brada{\'c} et al. (2002) are
compelling. The current claim is that in systems like B1422+231, 
the observed image magnification ratios cannot be obtained without
substructure.  The most general and convincing case was originally
presented by Schneider \& Mao (1998) for B1422+231 where there are 3
highly magnified images on one side of the lens centre (images A,B and
C), forming a line. On the other side of the lens lies a single,
fourth image D, that has a radio flux that is a factor of 50 lower
than that of the other three images. This means that either image D is
demagnified or the other 3 images are magnified very strongly. This,
however would imply that the source has to lie close to a cusp. There
exists a theorem that holds strictly only for sources on a cusp: the
sum of the flux of the outer two images ought to be equal to the flux
of the innermost image, or, written in terms of the magnifications:
\begin{equation}
\mu_{\rm A}+\mu_{\rm C}=\mu_{\rm B}\,\,{\rm for}\,\,\mu_{\rm tot}\rightarrow\infty,
\label{expected}
\end{equation}
where B is the central, brightest, image.  However, the images in
B1422+231 do not obey this relation.  Moving the source away from the
cusp would mean the theorem need not hold but implies that the total
magnification is small and hence that image D needs to be strongly
demagnified. Mao \& Schneider (1998) argue that this is unlikely as
the image is far from the lens center and hence in a low $\kappa$
region, leaving only the possibility that the source does lie close to
a cusp but that the magnification of one or more of the images is
effected by substructure in the vicinity.

Recently, Keeton et al. (2002) investigated the lensing properties of
sources lying in cusps in more detail.  His analysis is based on a
generic one-component elliptical power law lens model, including
external shear. He finds that the image magnification ratios of a
number of observed lens systems, B2045+265, RX J0911+0551 and
B0712+472 can not be fit by such "cusp lenses". No detailed modeling
is done in that paper and the argument is based on the deviant
magnification ratios alone.  The magnification ratio $(\mu_{\rm
A}+\mu_{\rm C})/\mu_{\rm B}$ of observed systems lies typically
between 1.5 and 2 for total magnifications of $\sim40$. One-component
lens models of these systems predict a magnification ratio $(\mu_{\rm
A}+\mu_{\rm C})/\mu_{\rm B}$ that is within a few per-cent of unity.  The
question arises whether the mismatch between observed and simulated
image magnification ratios is not due to the specific one-component
lens model used; a more complicated lens mass profile may explain the
observed magnification ratios.

As seen from the maps in Figs.~1-8 mergers and misalignments between
a halo and disk can cause a strong change in the magnification pattern
on the image plane. It is thus not unreasonable to propose that the
magnification ratios might be explained by a more complex structure in
the primary lens -- for example the presence of a misaligned disk, as
discussed in this paper.

In Figs.~9 and 10 we show the magnification ratio $1/|(\mu_{\rm
A}+\mu_{\rm C})/\mu_{\rm B}-1|$ for our Milky Way-type model and for
the early-type lens respectively for different offsets and disk
parameters. In all cases, only regions in the source plane that would
produce 4 detectable images are shown. In the case of the Fig.~9 this
is only true for regions in the source plane that form a total of 5
images. For the early-type lens model, regions that produce a total of
7 images in general also produce 4 bright images, the remaining three
images being strongly demagnified (cf. Fig~7, top-left and bottom-left
panels). To determine the magnification ratios for these systems we
proceed as follows: for each source position as defined by a grid on
the source plane consisting of $400\times400$ pixels of size
$2.5\times10^{-3}$arcsec, we identify the 4 most strongly magnified
images. Within this group, the image with the smallest magnification,
image D, is identified. Within the remaining group of three magnified
images we identify the central image as image B. The remaining two
images are assigned labels A and C randomly. Having thus identified
the images A,B,C and D for each source position, we determine the
magnification ratios from the individual magnifications as described
in \S~3.

The gray-scale in Figs.~9 and 10 is such that lightly-shaded regions
indicate source positions for which the magnification ratios do not
obey the cusp relation.  The top left panel in Fig.~8 shows the case
for no halo-disk misalignment for our Milky Way-type lens model; the disk in
the remaining three panels has a moderate misalignment in the x, the y
and the x-y direction.  The caustic shape changes significantly once
the disk is misaligned, leading to strong changes in the function of
magnification ratio vs. source position. Especially for a misalignment
in the y directory (top-left panel) and for misalignments of $\Delta
x=\Delta y=0.6"$, the caustic shape becomes very complicated and in
several cusp regions the expected magnification differs drastically
from that expected in "normal" cusps.  For the early-type lens, we
show the magnifcation ratio only for offsets of $\Delta x=\Delta
y=0.6"$, but for different disk parameters. The top left panel is for
the standard early-type lens model described in the previous
section. The three other panels are for three different sets of disk
masses and sizes. The bottom-right panel shows the results for a
small, light disk. In all cases sources lying in the upper-right cusp
would produce 3-1 image geometries with $(\mu_{\rm A}+\mu_{\rm
C})/\mu_{\rm B}$ magnification ratios that differ significantly from
the expected ratio of 1.

Note that for the Milky Way-type lens there is a strong effect due the
presence of the disk even when it is not misaligned. We show the
corresponding result for our standard early-type lens model in
Fig.~11. The scale is drastically different than that of Figs.~9 and 10;
the caustic structure is very small. Since we assume a spherical halo,
the asymmetry in the potential that causes the inner caustic is solely
due to the small disk embedded in the halo. However, there is not much
qualitative change compared to the MW-type lens: a disk affects the
magnification ratios of 4-image lens systems strongly also in E-type
lens galaxies. This is in concordance with the findings of M\"oller,
Hewett \& Blain who studied the effects of such aligned disks in
early-type lensing galaxies (MNRAS, submitted).

\section{Conclusions}

As has been illustrated in the plots in this paper, the presence of a
misaligned disky component embedded in a dark matter halo produces a
complex range of strong lensing effects - in the shapes and areas of
the caustics, magnification ratios, and image geometries.

Most importantly, we show that for models of high-magnification
($\mu_{\rm tot}\sim50$) systems the presence of a disky component can
substantially alter the expected magnification ratios and image
geometries. This is true for both Milky Way-type disks and small disks
as may be found in early-type galaxies. But misalignments of the disk
relative to the halo, such as might be caused by mergers or strong
interactions, can compound this effect.  The shape of the caustic
structure changes dramatically, and depending on the ellipticity of
the dark matter halo the enclosed area is enhanced by up to a factor
of 10. Thus, misaligned disks increase the overall cross-section for
the formation of four magnified images, while at the same time causing
strong deviation from the magnification ratios expected from simple
one-component lens models. Single component, elliptical mass models
are unable to reproduce such features, which may be the reason why
models based on simple elliptical mass models fail to reproduce the
observed image magnification ratios as for B1422+231.

Due to the complexity of the lens model investigated here we have
been concerned mainly with investigating the qualitative effect of
misaligned disks. Even though detailed modeling of existing systems
using a halo in combination with a misaligned disk lens is in
principle possible, this will be a complicated and time-consuming
task. Due to the large number of parameters and the relatively few
observational lensing constraints, it is very unlikely that a definite
model can be found; instead, it is to be expected that there will be a
considerable degree of degeneracy. However, our results do show
clearly that the simple one-component lens models currently used are
likely to be too simplistic and that more complicated models
including a possibly misaligned disk component can reproduce the observed
properties of many lens systems without introducing the need for halo
sub-clumps.

\acknowledgments

PN acknowledges a Research Fellowship from Trinity College, Cambridge.
OM acknowledges support from the Marie Curie Fellowship programme.

\begin{figure} 
\plotone{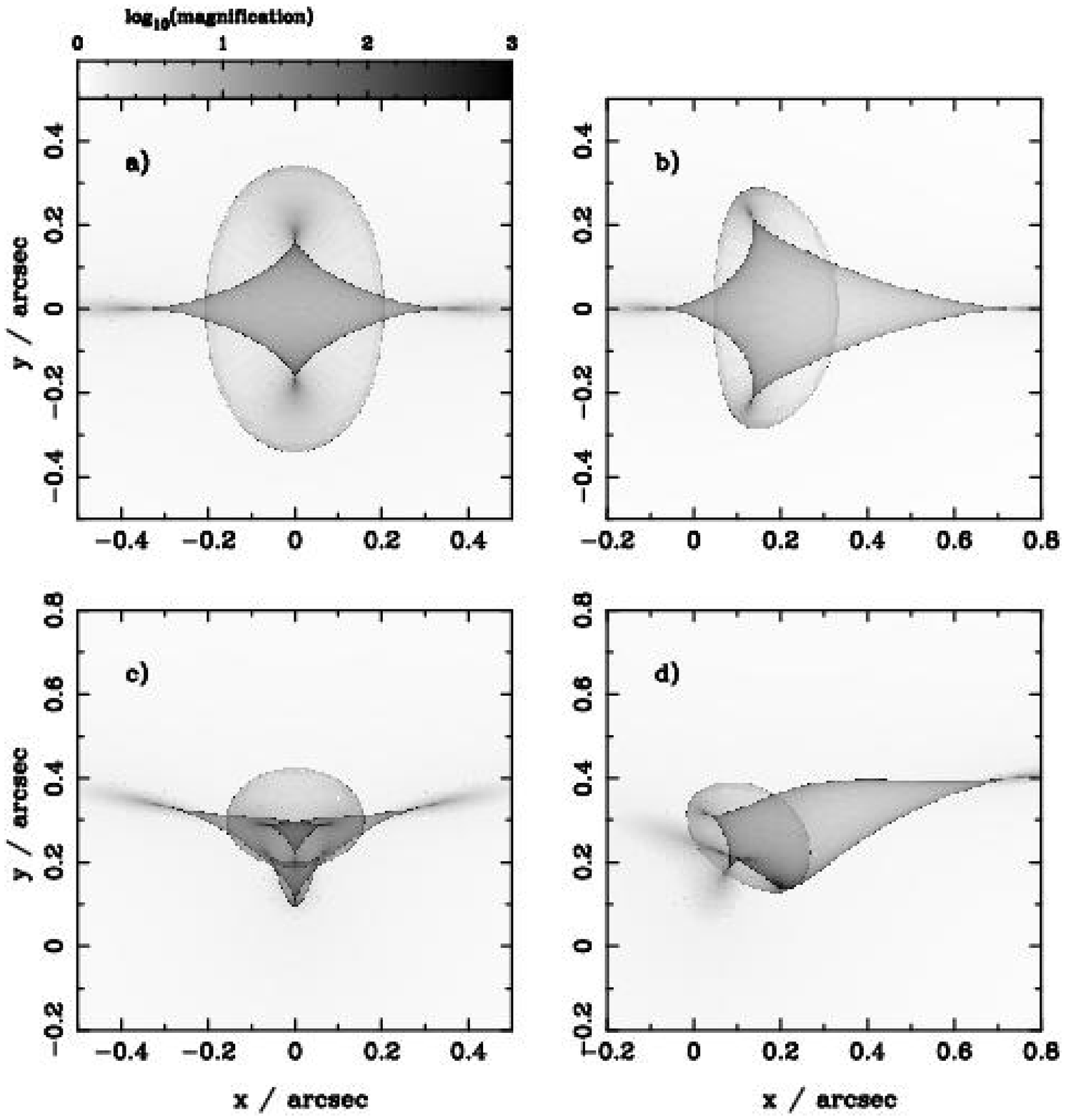}
\caption{The magnification maps in the source plane for the
spiral-lens system.  Spatial offsets of the disk are seen to strongly
distort both the diamond-shaped caustic and the elliptical
caustic. The location of the disk with respect to the center of halo
is [$\Delta x, \Delta y$]= (0'', 0'') [Panel a]; (0.6'', 0'') [Panel
b];(0'', 0.6'') [Panel c]; and (0.6'', 0.6'') [Panel d].  For our
choice of cosmological parameters and offset of 0.6'' corresponds to
2.3 kpc.}
\end{figure}

\begin{figure} 
\plotone{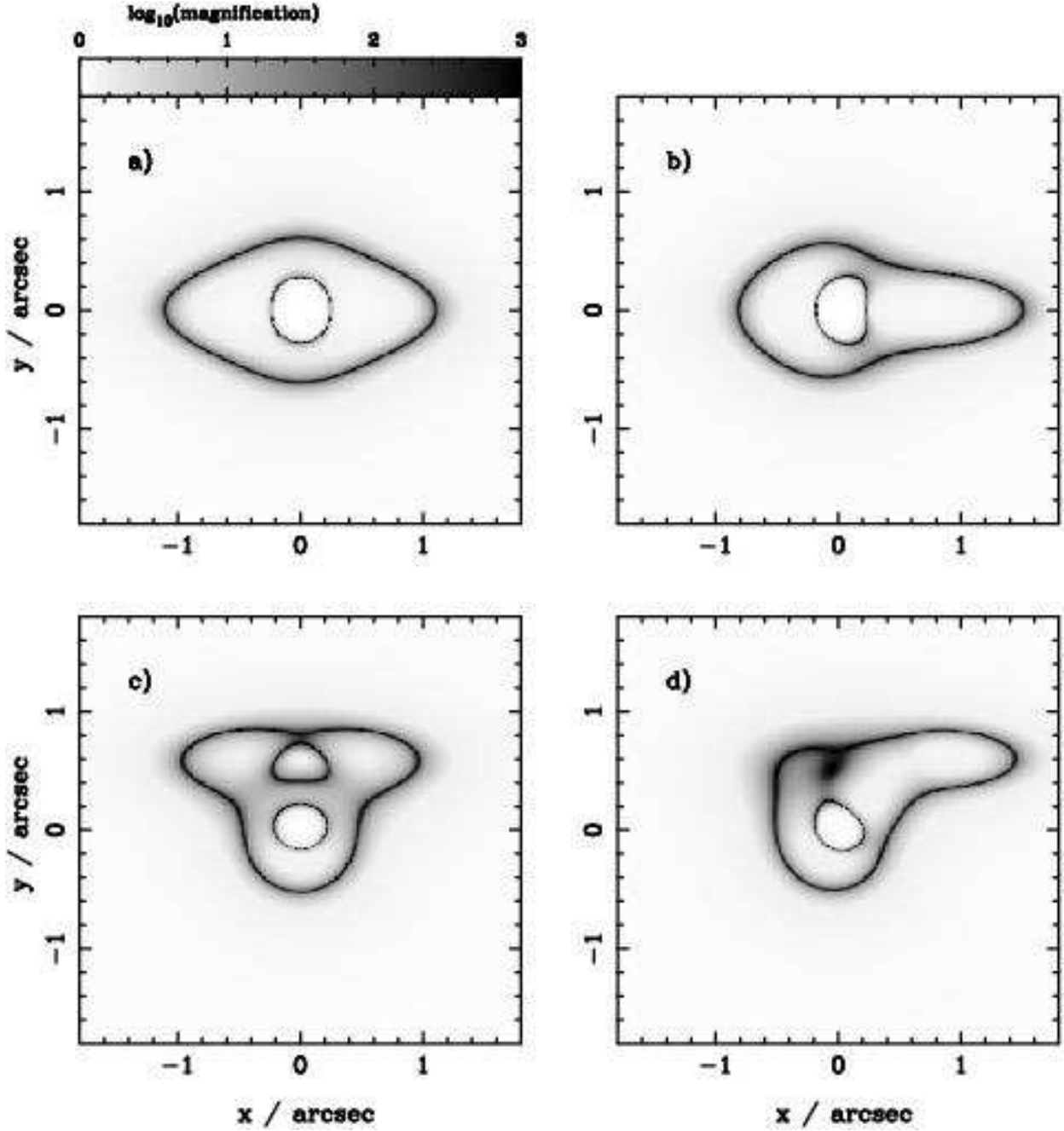}
\caption{Magnification maps in the image plane for the spiral lens.
The offsets are the same as those in Fig. 1.  The diamond-shaped
caustic in the source plane maps onto the outer critical curve in the
image plane, and the elliptical caustic maps onto the inner critical
curve.}
\end{figure}

\begin{figure} 
\plotone{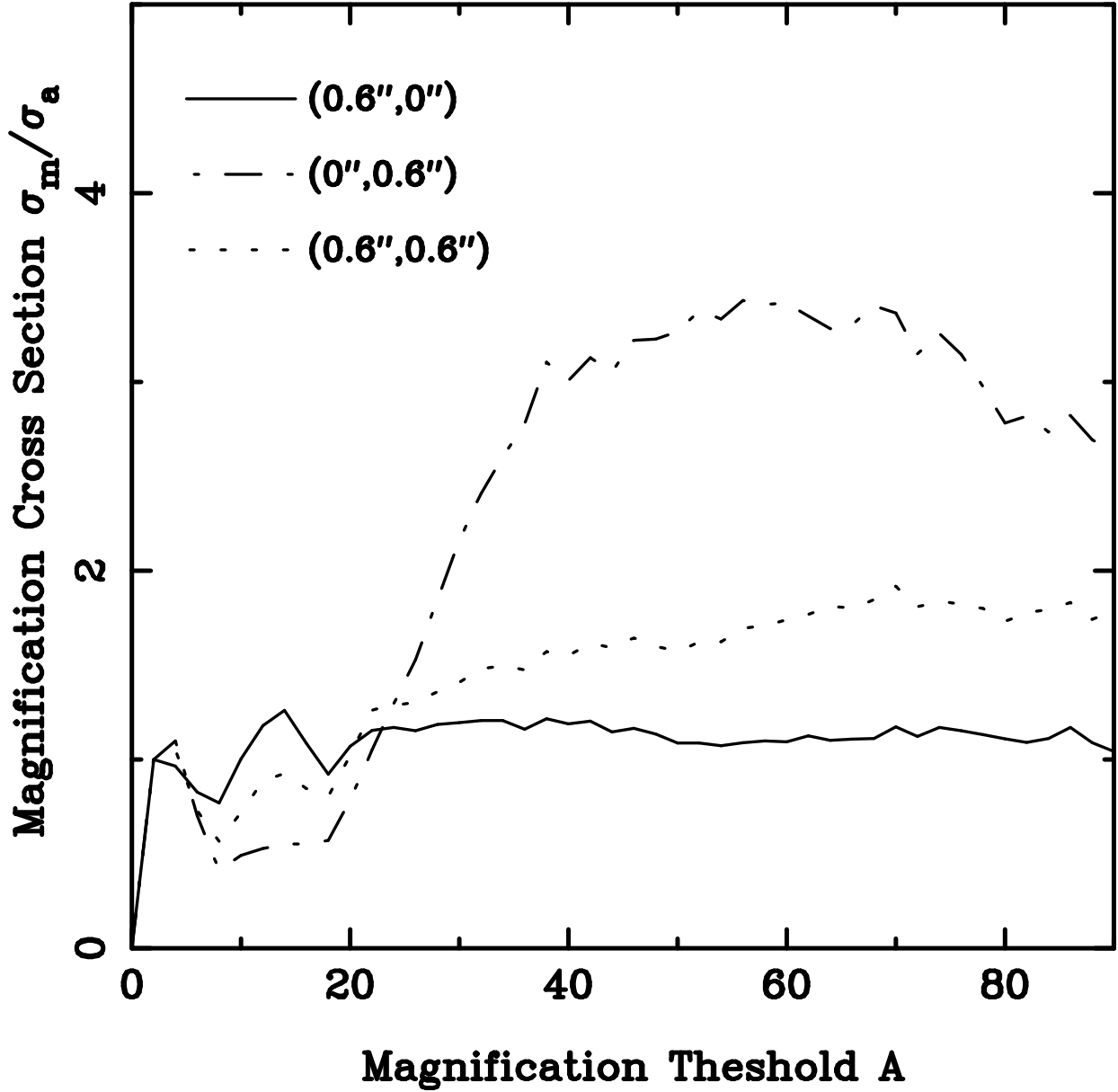}
\caption{The cross section for total magnification above a threshold
$A$ for the spiral lens.  The three curves specify the cross section
for our three misaligned cases relative to the aligned case.  The
differences become apparent at magnification ratios of $\sim$ 10 and
at the high magnification tail $>\,30$. Shifts in the y-direction (out of the
plane) have a much more significant effect than shifts in the x-direction.}
\end{figure}

\begin{figure} 
\plotone{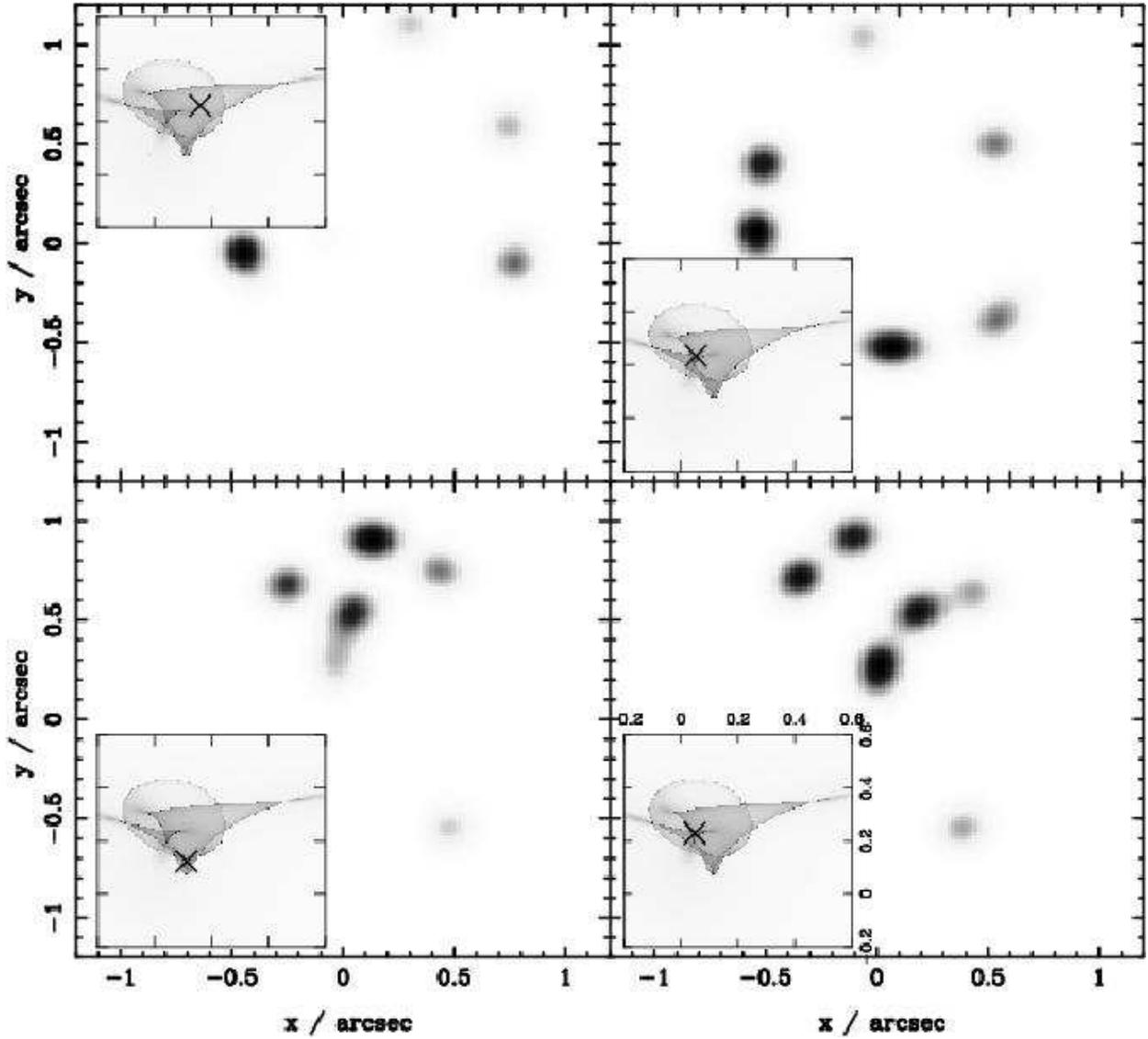}
\caption{Representative image geometries of a lensed point source and
a misaligned lens. The insets show the caustic geometry with the
location of the source marked by a cross. The four panels show the
resulting image configurations, which are more complex in the presence
of misaligned disks. The illustrative disk offset here is chosen to be
[$\Delta x , \Delta y$] = (0.3'', 0.6'')}
\end{figure}

\begin{figure} 
\plotone{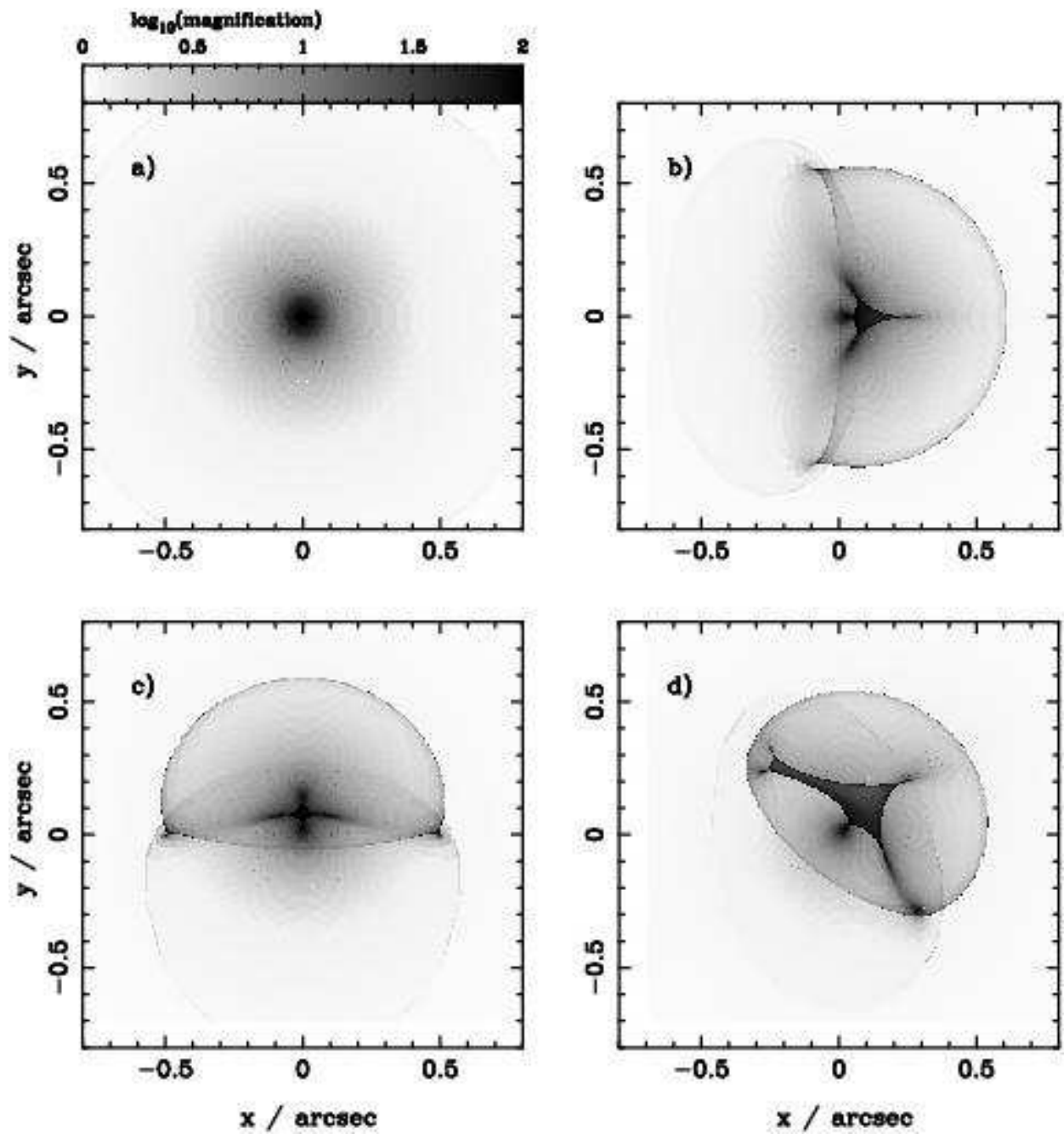}
\caption{Magnification maps in the source plane for various
lensing configurations of an early-type lens. The location of 
the disk with respect
to the center of the halo is [$\Delta x, \Delta y$] = (0'', 0'')
[Panel a]; (0.6'', 0'') [Panel b];(0'', 0.6'') [Panel c]; 
and (0.6'', 0.6'') [Panel d]. For our choice of cosmological parameters
an offset of 0.6'' corresponds to 2.3${\mathrm {kpc}}.$}
\end{figure}

\begin{figure}
\plotone{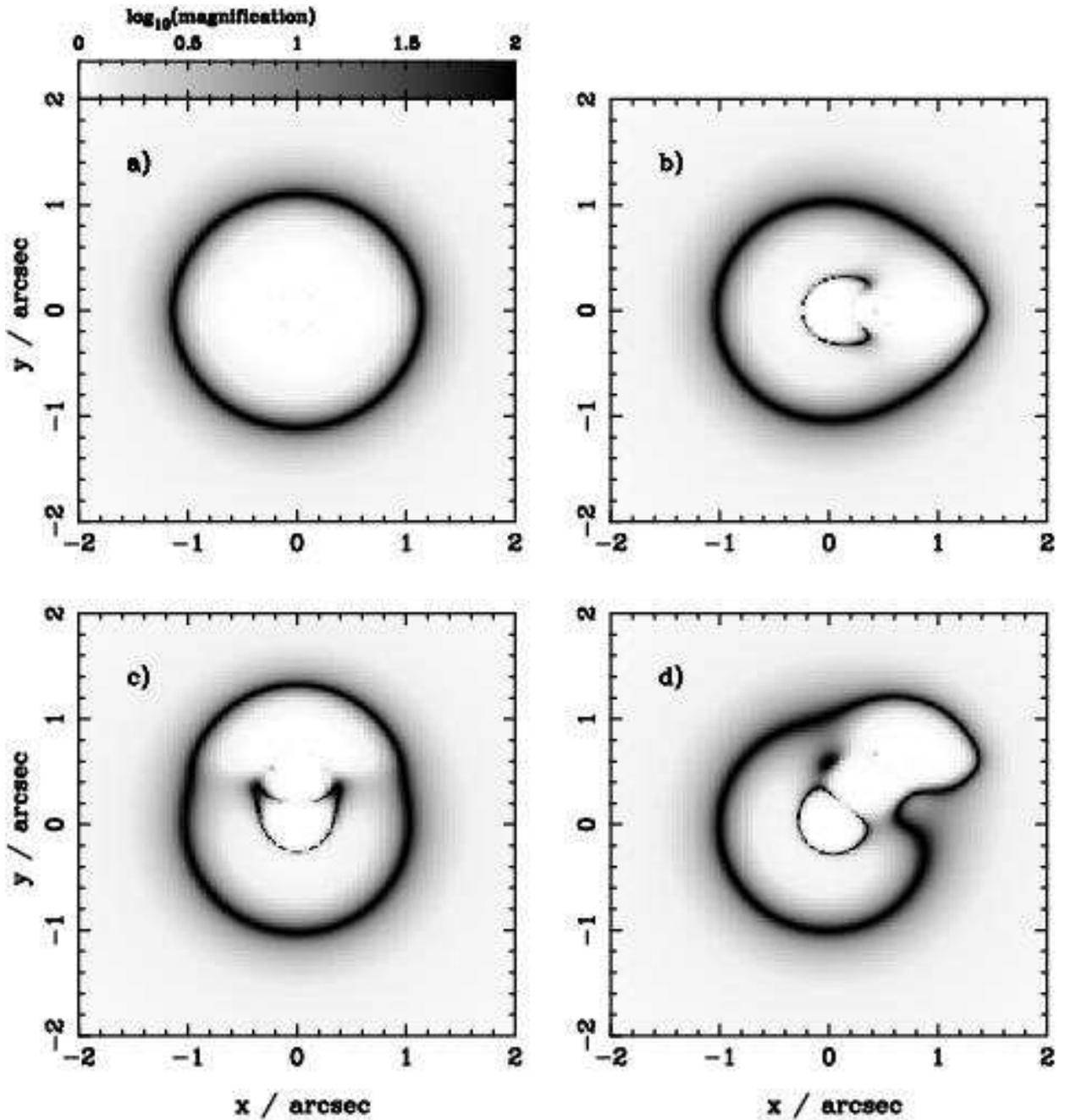}
\caption{Magnification maps in the image plane for the early-type lens.
The offsets are the same as those in Fig. 5.  Note that there is
additional structure within the outer critical curves that cannot be
seen.  For instance, there is a smooth diamond-like curve of low
magnification ($\sim$ 2) surrounding a highly de-magnified region at the
center of Panel (a).}
\end{figure}

\begin{figure} 
\plotone{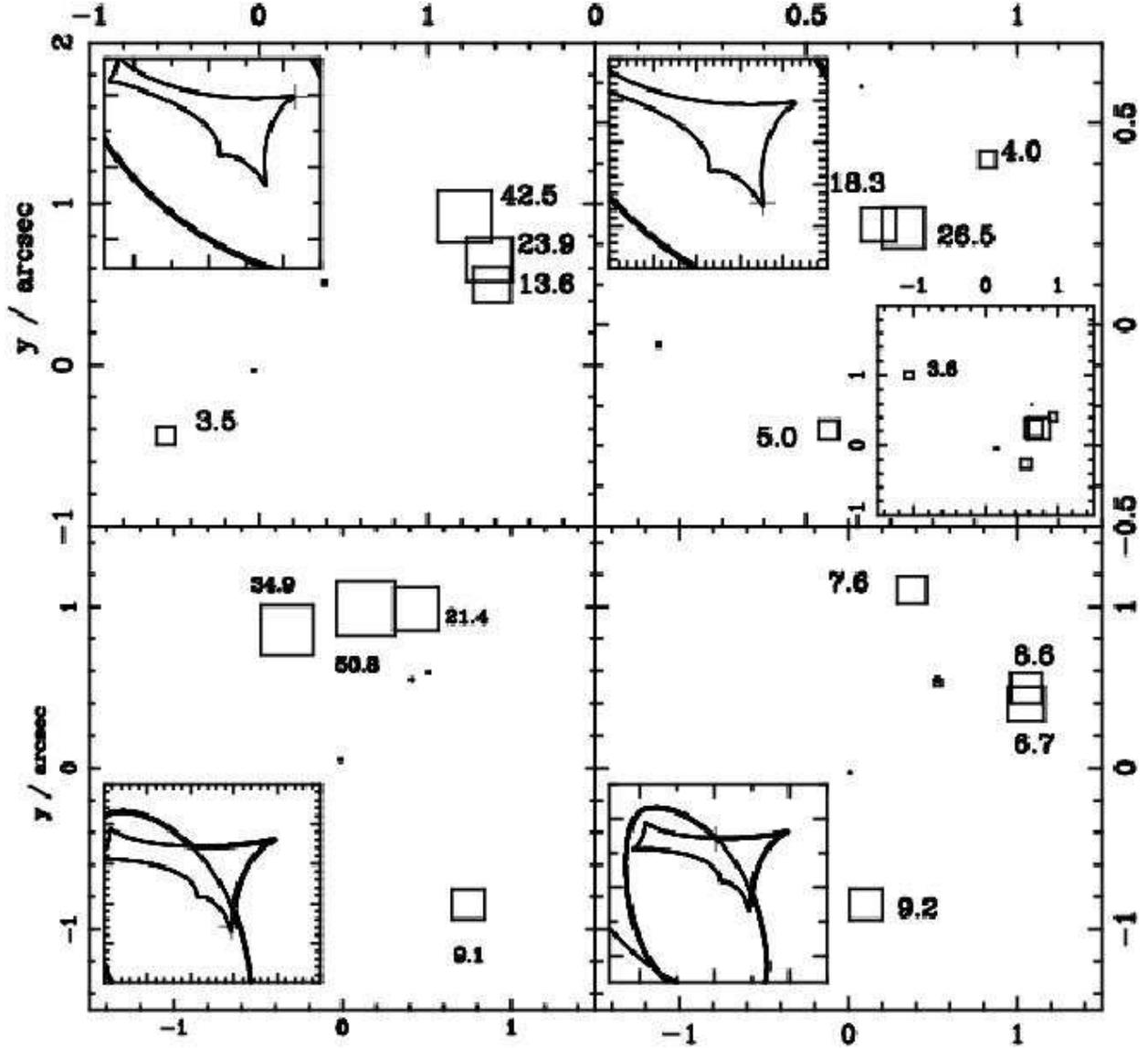}
\caption{Image geometries for two early-type lens models with a
misaligned disk. The disk has a mass of
$M_{\mathrm{d}}=2.0\times10^{10}M_{\odot}$ and a scale length of
$r_{\mathrm{d}}=0.5\,\mathrm{kpc}$ in the top two panels. In the
bottom two panels, the disk mass is
$M_{\mathrm{d}}=1.0\times10^{10}M_{\odot}$.  The other parameters are
the same as for Fig.~5, panel 3. In all panels, the boxes are centred
on the corresponding image positions and the size of the box is
proportional to the logarithm of the image magnification.  The numbers
next to the boxes give the image magnifications. The small boxes
without numbers indicate the positions of demagnified images.  In each
panel, the inset shows the central $0.6\arcsec\times0.6arcsec$ region
around the lens centre, the lines marking the caustic lines. The cross
in each inset marks the position of the point source, which is
different in each panel.  The scale in the top left panel is smaller
and does not cover one of the images. A second subpanel shows all the
images on a larger scale.}
\end{figure}

\begin{figure} 
\plotone{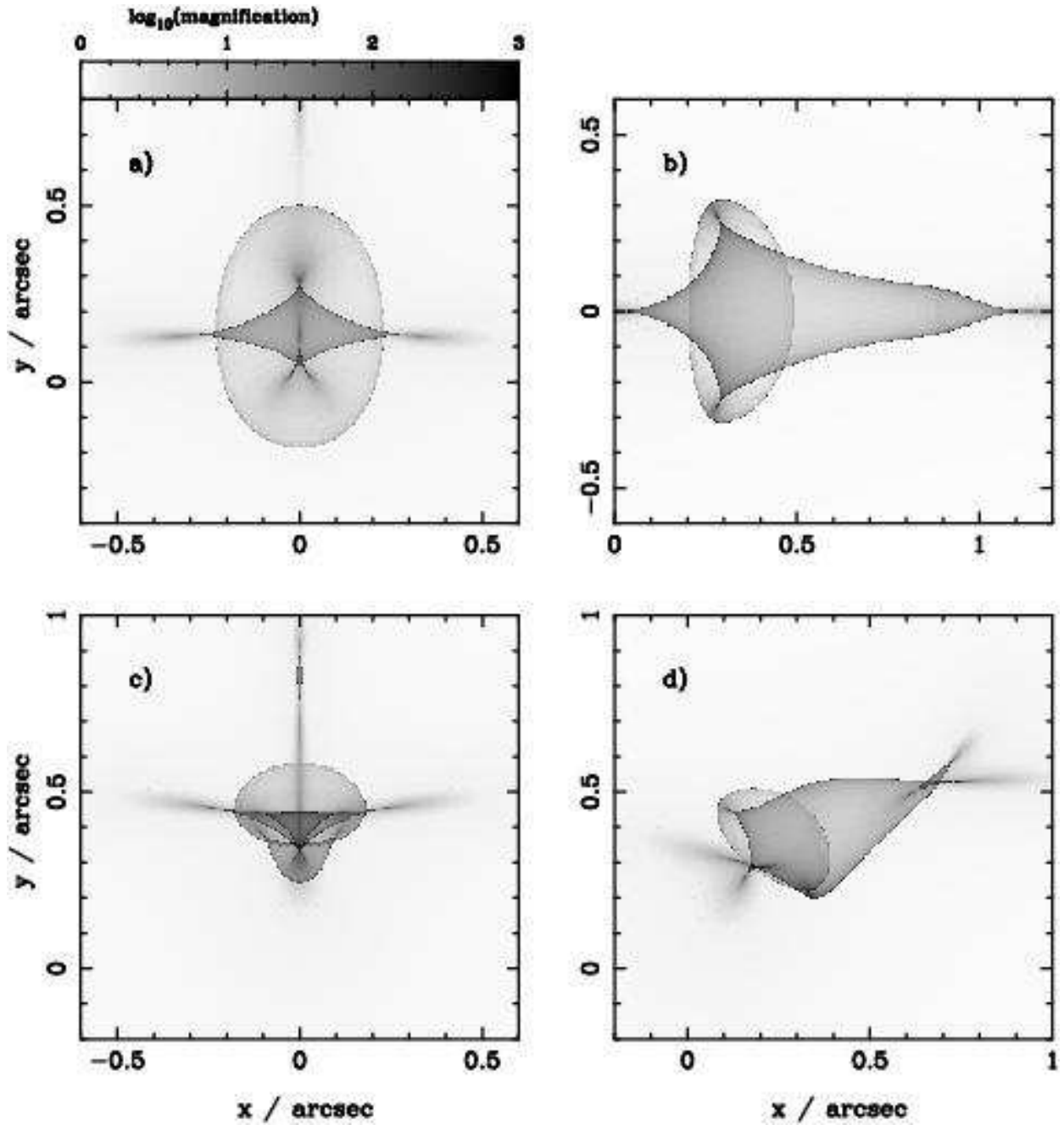}
\caption{Magnification maps in the source plane for interacting
halos. The secondary halo is located roughly 1.7'' away from the
primary lens in the same direction as the disk offset, except in the
first Panel where it is aligned along the y-axis. The offsets of the
disks with respect to the dark halos are once again as in Fig.~1.}
\end{figure}

\begin{figure}
\plotone{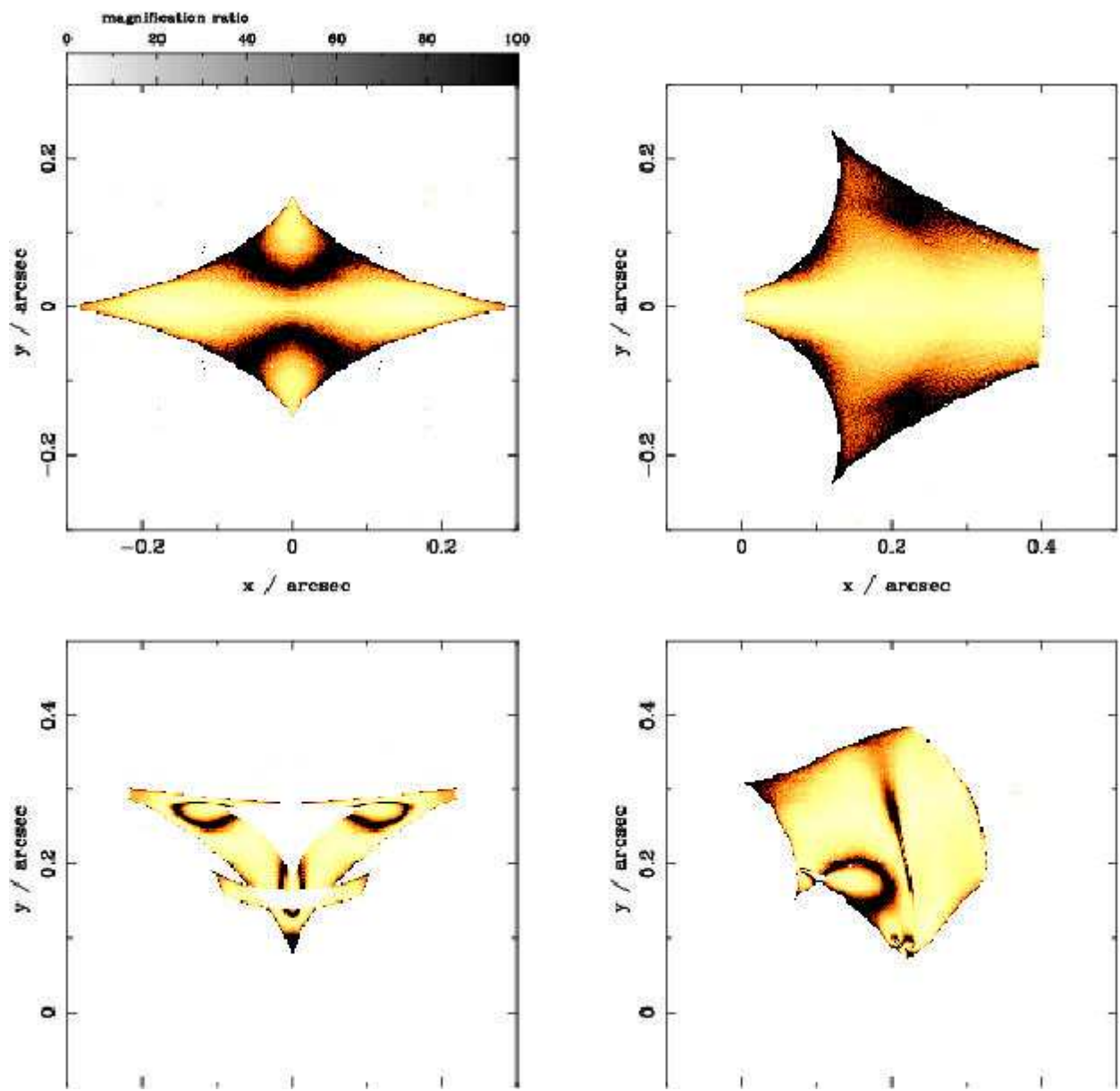}
\caption{The magnification ratio $1/|(\mu_{\rm A}+\mu_{\rm
C})/\mu_{\rm B}-1|$ for all four image systems for the Milky Way type
model. The three misaligned cases studied are shown together with the
aligned case for comparison. The gray-scale gives an indication of how
much the magnification ratios of the 4 image systems deviate from
those expected in the limit of infinite total magnification. A lighter
shade indicates a greater deviation with the magnification ratios
expected for $\mu_{\rm tot}\rightarrow\infty$. Note that even in
regions very close to the caustic, where $\mu_{\rm tot}\sim100$, there
may be large deviations from the expected magnification ratios.}
\end{figure}

\begin{figure} 
\plotone{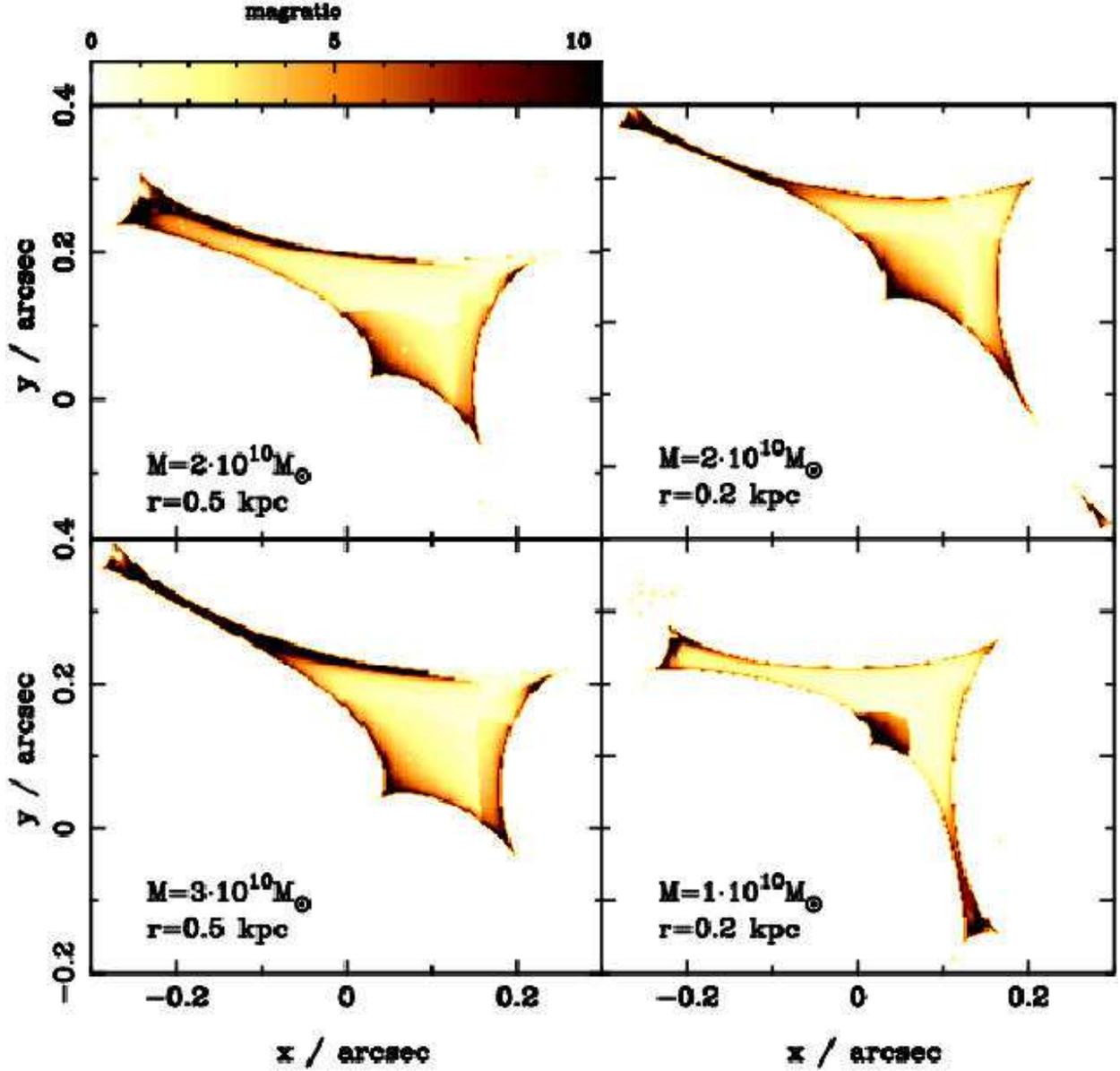}
\caption{The magnification ratio $1/|(\mu_{\rm A}+\mu_{\rm
C})/\mu_{\rm B}-1|$ for four elliptical lens models with misaligned
disks. In all cases, the offset of the disk is $\Delta x=\Delta
y=0.6"$. The halo is modelled as a CIS with velocity dispersion of
$\sigma_{\mathrm{v}}=210\mathrm{km}{s^{-1}}$ and core radius
$r_{\mathrm{c}}=0.5\mathrm{kpc}$ in all four panels. The exponential
disk parameters differ between the panels; disk scale lengths and
masses are indicated in the bottom left of each panel. The parameters
of the disks are representative to those obtained from bulge+disk
light profile fits to ellipticals in CL1358+62 (Kelson et al., 2000).
The gray-scale is as in the previous figure; a lighter shade indicates
a greater deviation with the magnification ratios expected for
$\mu_{\rm tot}\rightarrow\infty$.}
\end{figure}

\begin{figure} 
\plotone{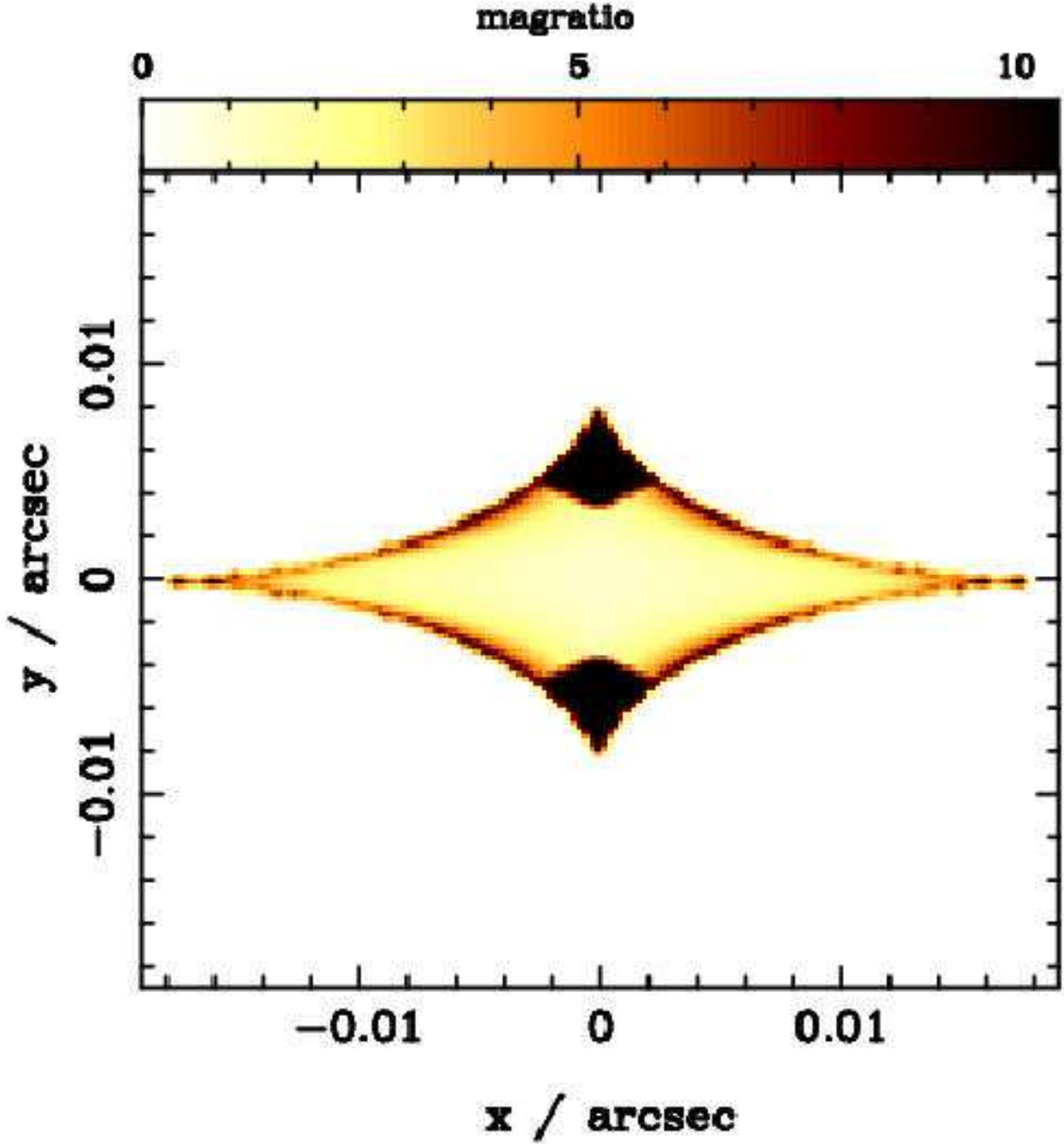}
\caption{The magnification ratio $1/|(\mu_{\rm A}+\mu_{\rm
C})/\mu_{\rm B}-1|$ for an elliptical galaxy with aligned disk. The
parameters are as in Fig~10, top left panel, but $\Delta x=\Delta
y=0$. The gray-scale is as in the previous figures. Note the much
smaller scale as compared to Figs.~9 and 10.}
\end{figure}

\end{document}